\documentclass[preprint,showpacs,amsmath,amssymb,superscriptaddress]{revtex4}
\usepackage{amsmath,amssymb,color,latexsym,graphicx}
\usepackage{dcolumn}
\usepackage{bm}
\usepackage{color}

\def\be{\begin{equation}}
\def\ee{\end{equation}}
\def\ba{\begin{eqnarray}}
\def\ea{\end{eqnarray}}
\def\epa{\boldsymbol{e}_\parallel}
\def\kpn{k_{\perp}}
\def\ppn{p_{\perp}}
\def\qpn{q_{\perp}}
\def\kpa{k_{\parallel}}
\def\kk{\boldsymbol{k}}
\def\pp{\boldsymbol{p}}
\def\qq{\boldsymbol{q}}
\def\kknb{\boldsymbol{k_{\perp}}}
\def\ppnb{\boldsymbol{p_{\perp}}}
\def\qqnb{\boldsymbol{q_{\perp}}}
\def\uu{{\bf u}}
\def\ww{{\boldsymbol w}}
\def\ome{{\boldsymbol \Omega}_0}

\def \pmbtext#1{\leavevmode
     \setbox0\hbox{#1}
     \kern-0,2pt \copy0 \kern-\wd0
     \kern0,4pt \copy0 \kern-\wd0
     \kern-0,2pt \raise0,3pt \box0 }

\def\bnabla{{\pmb \nabla}}

\begin{document}

\title{A multiple time scale approach for anisotropic inertial wave turbulence}

\author{S\'ebastien Galtier}
\affiliation{Universit\'e Paris-Saclay, Institut universitaire de France, 
Laboratoire de Physique des Plasmas, Ecole polytechnique, 91128 Palaiseau, France}

\begin{abstract}
Wave turbulence is the study of the long-time statistical behaviour of equations describing a set of weakly non-linear interacting waves. Such a theory, which has a natural asymptotic closure, allows us to probe the nature of turbulence more deeply than the exact Kolmogorov laws by rigorously proving the direction of the cascade and the existence of an inertial range, predicting stationary spectra for conserved quantities, or evaluating the Kolmogorov constant. An emblematic example is given by fast rotating fluids for which a wave turbulence theory has been derived by \citet{Galtier2003}. This work involves nontrivial analytical developments for a problem that is anisotropic by nature. We propose here a new path for the derivation of the kinetic equation by using the anisotropy at the beginning of the analysis. We show that the helicity basis is not necessary to obtain the wave amplitude equation for the canonical variables that involve a combination of poloidal and toroidal fields. The multiple time scale method adapted to this anisotropic problem is then used to derive the kinetic equation which is the same as the original work when anisotropy is eventually taken into account. This result proves the commutativity between asymptotic closure and anisotropy. In addition, the multiple time scale method informs us that the kinetic equation can be derived without imposing restrictions on the probability distribution of the wave amplitude such as quasi-Gaussianity, or on the phase such as random phase approximation which naturally occurs dynamically.
\end{abstract}

\maketitle

\section{Introduction}
As we celebrate two hundred years of the Navier-Stokes equations, it is remarkable to note that hydrodynamic turbulence is still a much studied subject. Among the different fields of study, there is wave turbulence which has become important in geophysics and astrophysics where waves are omnipresent \cite{GaltierCUP2023}. 
The strength of the (weak) wave turbulence theory is that it offers the possibility of a deep understanding of physical systems composed of a set of random waves interacting nonlinearly. The reason for this is, first of all, the possibility of analytically deriving a set of integro-differential equations for second-order spectral cumulants -- the so-called kinetic equations -- which are free from the closure problem classically encountered in eddy turbulence. 
Indeed, in wave turbulence there is a natural asymptotic closure rooted on the existence of a small parameter, the wave amplitude.
Secondly, exact solutions (Kolmogorov-Zakharov spectra) can be found from the kinetic equations. In addition to the usual thermodynamic solutions, the kinetic equations have finite flux solutions that capture the flow of conserved densities from sources to sinks. Thirdly, these solutions correspond to power law spectra that can be compared to the data. The number of experiments, observations, and diagnostics has increased considerably over the past two decades and today, thanks also to direct numerical simulations (DNS), wave turbulence has become a leading field in turbulence where new fundamental questions are being raised \citep{Galtier2017,Hassaini2017,Hassaini2019,Savaro2020,Galtier2021,Ricard2021,David2022,Falcon2022,Griffin2022,Hrabski2022,Kochurin2022,Onorato2022,Rodda2022,Zhang2022,Dematteis2023,Galtier2023,Lanchon2023,Novkoski2023,Zhu2023}. 

Rotating fluids are one of the most studied examples in (strong/weak) wave turbulence as it involves inertial waves that are easy to excite experimentally \citep{Hopfinger1982,Jacquin1990,Morize2005}. This regime is of interest in a number of fields such as geophysics where the Coriolis force is felt, for example, through large-scale atmospheric motions. Very early, it was recognised that rotating turbulence behaves differently from classical eddy turbulence with a reduction of the cascade along the axis of rotation ${\bf \Omega_0}$ with possibly a steeper energy spectrum than the well-known Kolmogorov spectrum \citep{Hossain1994,Zeman1994,Zhou1995,Cambon1997,Smith1999,Baroud2002,Godeferd2015}. Another remarkable feature, still not understood, is the self-similarity found in the scaling of velocity structure functions in the direction transverse to $\ome = \Omega_0 \epa$ \citep{Baroud2002,vanBokhoven2009}, which is in strong contrast with intermittency observed in hydrodynamic turbulence. 
In recent years, the (weak) inertial wave turbulence regime has been specifically studied experimentally \citep{Yarom2014,Monsalve2020}. For example, it was shown that the energy spectrum is concentrated along the dispersion relation, as expected in wave turbulence, with a scaling in agreement with the theoretical prediction. Numerical simulations (including DNS) have also been carried out, notably to study the spectral properties 
\citep{Bellet2006,Scott2014,Clark2016,LeReun2017,Sharma2018,Galtier2020b,LeReun2020,Yokoyama2021}. They confirm the previous experimental and theoretical results and reveal, for example, the existence of a non-stationary solution different from the stationary Kolmogorov-Zakharov spectrum. 

The theory of inertial wave turbulence has been derived by \citet{Galtier2003}. The kinetic equation obtained is valid in the most general case, i.e. without making the assumption of anisotropy. However, a simple argument based on the resonance condition shows that the cascade is anisotropic with a transfer mainly in the perpendicular ($\perp$) direction to $\ome$. Using this feature, the kinetic equation was eventually reduced to the axisymmetric case from which the exact (Kolmogorov-Zakharov) energy spectrum was derived. This solution takes the form $E(\kpn,\kpa) \sim \kpn^{-5/2} \kpa^{-1/2}$, with $\kk = \kk_{\perp} + \kpa \epa$ the wavevector. As recently proved by \citet{David2023}, this energy spectrum corresponds to a local turbulence (with an inertial range independent of the largest and smallest scale) for which we can also estimate the Kolmogorov constant. The derivation of the kinetic equation of inertial wave turbulence in the general case (without the axisymmetry assumption) is cumbersome and the use of the Hamiltonian formalism does not drastically simplify the calculation \citep{Gelash2017}. 
Here, it is proposed to take another path for the derivation of such an equation by using the anisotropy assumption ($\kpn \gg \kpa$) at the beginning of the analysis. As shown in \S \ref{section2}, in this case the wave amplitude equation can be obtained without the introduction of a complex helicity basis, which is an interesting simplification: the velocity field is decomposed into poloidal and toroidal fields from which we can define the canonical variables. In \S \ref{section3}, the multiple time scale method introduced by \citet{Benney1966} is adapted (a few points are also clarified) to this anisotropic problem and then used to derive such a kinetic equation which is the same as the original work when anisotropy is finally taken into account. This result shows the commutativity between asymptotic closure and anisotropy. Furthermore, the multiple time scale method informs us that the kinetic equation 
(of weak wave turbulence)
can be derived without imposing restrictions on the probability distribution of the wave amplitude such as quasi-Gaussianity, or on the phase such as random phase approximation which naturally occurs dynamically. Finally, we conclude with a discussion in \S \ref{section4},

\section{Wave amplitude equation}
\label{section2}

\subsection{Canonical variables}
The Navier-Stokes equations with the Coriolis force read
\be
\frac{\partial \ww}{\partial t} - 2 (\ome \cdot \bnabla) \uu = (\ww \cdot \bnabla)  \uu - (\uu \cdot \bnabla) \ww + \nu \, \nabla^2  \ww , \label{NSrot}
\ee
where $\uu$ is a solenoidal velocity ($\bnabla \cdot \uu = 0$), $\ww = \bnabla \times \uu$ the vorticity and $\ome$ a constant rotation rate. Hereafter, we will neglect the viscosity. We introduce the toroidal ($\psi$) and poloidal ($\phi$) scalar fields in the following manner
\be
\uu = \nabla \times (\psi \epa) + \nabla \times (\nabla \times (\phi \epa) ) ,
\ee
whose Fourier transform writes
\be
\hat \uu_k = i \hat \psi_k \kk \times \epa -  \hat \phi_k \kk \times (\kk \times \epa ) 
= i \hat \psi_k \kk \times \epa + \hat \phi_k ( k^2 \epa - \kpa \kk ) ,
\ee
from which we deduce the vorticity vector ($\vert \kk \vert =k$)
\be
\hat \ww_k = \hat \psi_k ( k^2 \epa- \kpa \kk) + i k^2 \hat \phi_k  \kk \times \epa . 
\ee
It is straightforward to show in Fourier space that the linear contribution of equation (\ref{NSrot}) leads, after projection, to 
\begin{subequations}
\ba
\frac{\partial \hat \phi_k}{\partial t} &=& 2i \Omega_0 \frac{\kpa}{k^2} \hat \psi_k , \\
\frac{\partial \hat \psi_k}{\partial t} &=& 2i \Omega_0 \kpa \hat \phi_k .
\ea
\end{subequations}
The linear solutions are the well-known (helical) inertial waves with the angular frequency ($\partial^2_t = - \omega^2_k$ can be used)
\be
\omega^2_k = 4 \Omega^2_0 \frac{\kpa^2}{k^2} .
\ee
From this property, we introduce the canonical variables 
\be \label{vcanon}
A^{s}_k \equiv A^{s} (\kk) = k^2 \hat \phi_k - s k \hat \psi_k ,
\ee
with $s=\pm$ the directional polarity. With such a choice of canonical variables, we have
\be
\vert A^+_k \vert^2 + \vert A^-_k \vert^2 = 2 \vert \hat \uu_k \vert^2 
\ee
and at the linear level
\be
\frac{\partial A^s_k}{\partial t} + i s \omega_k A^s_k = 0 .
\ee

\subsection{Resonance condition}
The resonance condition for three-wave interactions can be written \citep{GaltierCUP2023}
\begin{subequations}
\ba
s \omega_k + s_p \omega_p + s_q \omega_q &=& 0, \\
\kk + \pp + \qq &=& 0 .
\ea
\end{subequations}
In the case of inertial waves, these relations are equivalent to the conditions
\be \label{reson}
\frac{s_q q - s_p p}{s \omega_k} = \frac{sk - s_q q}{s_p \omega_p} = \frac{s_p p - sk}{s_q \omega_q} . 
\ee
Assuming that the system is initially excited at large scale in a narrow isotropic domain in Fourier space, a situation often considered in DNS, the dynamics will initially be dominated by local interactions such that $k \simeq p \simeq q$. As the locality of the interactions is a property of turbulence that is generally verified, we can extend its use beyond the initial instant. We obtain
\be
\frac{s_q -s_p}{s \kpa} \simeq \frac{s - s_q}{s_p p_\parallel} \simeq \frac{s_p - s}{s_q q_\parallel} . 
\ee
From this expression, we can show that the associated cascade is necessarily anisotropic. Indeed, if $\kpa$ is non-zero, the left-hand term will only give a non-negligible contribution when $s_{p}=-s_{q}$. The immediate consequence is that either the middle or the right-hand term has its numerator which cancels (to leading order), which implies that the associated denominator must also cancel (to leading order) to satisfy the equality: for example, if $s=s_{p}$ then $q_\parallel \simeq 0$. This condition means that the transfer in the parallel direction is negligible: indeed, the integration of the wave amplitude equation in the parallel direction (see below) is then reduced to a few modes (since $p_\parallel \simeq \kpa$) which strongly limits the transfer between parallel modes. The cascade in the parallel direction is thus possible but relatively weak compared to that in the perpendicular direction. In the following, we will take advantage of this property and consider the anisotropic limit $\kpn \gg \kpa$ to simplify the derivation. 
Note that once turbulence is anisotropic, we can still use the locality condition with $k \sim k_\perp$; we then obtain $k_\perp \sim p_\perp \sim q_\perp$, whereas the parallel wavenumbers are limited to a narrow domain.

\subsection{Wave amplitude equation}

In the derivation of the wave amplitude equation, we will consider a continuous medium which can lead to mathematical difficulties connected with  infinite dimensional phase spaces. For this reason, it is preferable to assume a variable spatially periodic over a box of finite size $L$. However, in the derivation of the kinetic equation, the limit $L \to +\infty$ is finally taken (before the long time limit, or equivalently the limit $\epsilon \to 0$). As both approaches lead to the same kinetic equation, for simplicity, we anticipate this result and follow the original approach of \citet{Benney1966}. 
Note that the anisotropic limit ($\kpn \gg \kpa$) will also be taken before the (asymptotic) long time limit. 

The first non-linear term of equation (\ref{NSrot}) writes
\ba
\widehat{(\ww \cdot \bnabla) \uu}_k &=& i \int (\hat \ww_p \cdot \qq) \hat \uu_q \delta_{k,pq} d\pp d\qq \nonumber \\
&=& i \int \left[ i \hat \phi_p \hat \phi_q p^2 \left( \qq \cdot (\pp \times \epa)\right) (q^2 \epa -q_\parallel \qq) 
- \hat \phi_p \hat \psi_q p^2 \left( \qq \cdot (\pp \times \epa)\right) (\qq \times \epa) \right. \nonumber \\
&&\left. \mbox{} - \hat \psi_p \hat \phi_q \left( p_\parallel \pp \cdot \qq - p^2 q_\parallel \right) (q^2 \epa -q_\parallel \qq) 
- i \hat \psi_p \hat \psi_q \left( p_\parallel \pp \cdot \qq - p^2 q_\parallel \right) (\qq \times \epa)  \right] \nonumber \\
&& \mbox{}\times \delta_{k,pq} d\pp d\qq ,
\ea
with $\delta_{k,pq} \equiv \delta(\kk-\pp-\qq)$. In the anisotropic limit ($\kpn \gg \kpa$), a first simplification arises
\ba
\widehat{(\ww \cdot \bnabla) \uu}_k &=& i \int \left[ i \hat \phi_p \hat \phi_q \ppn^2 \qpn^2 \left( \epa \cdot (\qqnb \times \ppnb)\right) \epa
- \hat \phi_p \hat \psi_q \ppn^2 \left( \epa \cdot (\qqnb \times \ppnb)\right) (\qqnb \times \epa) \right. \nonumber \\
&&\left. \mbox{} - \hat \psi_p \hat \phi_q \qpn^2 \left( p_\parallel \ppnb \cdot \qqnb - \ppn^2 q_\parallel \right) \epa 
- i \hat \psi_p \hat \psi_q \left( p_\parallel \ppnb \cdot \qqnb - \ppn^2 q_\parallel \right) (\qqnb \times \epa)  \right] \nonumber \\
&& \mbox{} \times \delta_{k,pq} d\pp d\qq .
\ea

The second non-linear term of equation (\ref{NSrot}) reads
\ba
\widehat{(\uu \cdot \bnabla) \ww}_k &=& i \int (\hat \uu_p \cdot \qq) \hat \ww_q \delta_{k,pq} d\pp d\qq \nonumber \\
&=& i \int \left[ i \hat \phi_p \hat \phi_q \left( p^2 q_\parallel -p_\parallel \pp \cdot \qq \right) q^2 (\qq \times \epa) 
+ \hat \phi_p \hat \psi_q \left( p^2 q_\parallel - p_\parallel \pp \cdot \qq \right) (q^2 \epa -q_\parallel \qq) \right. \nonumber \\
&&\left. \mbox{} - \hat \psi_p \hat \phi_q \left( \qq \cdot (\pp \times \epa)  \right) q^2 (\qq \times \epa)
+ i \hat \psi_p \hat \psi_q \left( \qq \cdot (\pp \times \epa) \right) (q^2 \epa - q_\parallel \qq) \right] \nonumber \\
&& \mbox{} \times \delta_{k,pq} d\pp d\qq ,
\ea
which simplifies in the anisotropic limit to
\ba
\widehat{(\uu \cdot \bnabla) \ww}_k &=& i \int \qpn^2 \left[ i \hat \phi_p \hat \phi_q \left( \ppn^2 q_\parallel -p_\parallel \ppnb \cdot \qqnb \right) (\qqnb \times \epa) 
+ \hat \phi_p \hat \psi_q \left( \ppn^2 q_\parallel - p_\parallel \ppnb \cdot \qqnb \right) \epa \right. \nonumber \\
&&\left. \mbox{} - \hat \psi_p \hat \phi_q \left( \epa \cdot (\qqnb \times \ppnb) \right) (\qqnb \times \epa)
+ i \hat \psi_p \hat \psi_q \left( \epa \cdot (\qqnb \times \ppnb) \right) \epa \right] \nonumber \\
&& \mbox{} \times \delta_{k,pq} d\pp d\qq .
\ea
The addition of these two non-linear contributions leads to the simplified expression
\ba
\widehat{NL}(\kk) &=& \widehat{(\ww \cdot \bnabla) \uu}_k - \widehat{(\uu \cdot \bnabla) \ww}_k \nonumber \\
&=&\int \hat \phi_p \hat \phi_q \ppn^2\qpn^2 \left( \epa \cdot (\ppnb \times \qqnb)\right) \epa \delta_{k,pq} d\pp d\qq \nonumber \\
&& \mbox{} +i \int \hat \phi_p \hat \psi_q \ppn^2 \left( \epa \cdot (\ppnb \times \qqnb)\right) (\qqnb \times \epa) \delta_{k,pq} d\pp d\qq  \nonumber \\
&& \mbox{} -i \int \hat \psi_p \hat \phi_q \qpn^2 \left( \epa \cdot (\ppnb \times \qqnb) \right) (\qqnb \times \epa) \delta_{k,pq} d\pp d\qq  \nonumber \\
&& \mbox{} - \int \hat \psi_p \hat \psi_q \qpn^2 \left( \epa \cdot (\ppnb \times \qqnb) \right) \epa \delta_{k,pq} d\pp d\qq . 
\ea
The introduction of the canonical variables 
\begin{subequations}
\ba
\hat \psi_k &=& - \frac{1}{2 \kpn} \sum_s s A_k^s , \\
\hat \phi_k &=& \frac{1}{2 \kpn^2} \sum_s A_k^s ,
\ea
\end{subequations}
gives
\ba
\widehat{NL}(\kk) &=& \frac{1}{4} \sum_{s_p s_q} \int A_p^{s_p} A_q^{s_q} \left( \epa \cdot (\ppnb \times \qqnb)\right) \epa \delta_{k,pq} d\pp d\qq \nonumber \\
&&\mbox{} -\frac{i}{4} \sum_{s_p s_q} \int A_p^{s_p} A_q^{s_q} \frac{s_q}{\qpn} \left( \epa \cdot (\ppnb \times \qqnb)\right) (\qqnb \times \epa) \delta_{k,pq} d\pp d\qq  \nonumber \\
&&\mbox{}  + \frac{i}{4} \sum_{s_p s_q} \int A_p^{s_p} A_q^{s_q} \frac{s_p}{\ppn} \left( \epa \cdot (\ppnb \times \qqnb) \right) (\qqnb \times \epa) \delta_{k,pq} d\pp d\qq  \nonumber \\
&&\mbox{}  - \frac{1}{4} \sum_{s_p s_q} \int A_p^{s_p} A_q^{s_q} s_p s_q  \frac{\qpn}{\ppn} \left( \epa \cdot (\ppnb \times \qqnb) \right) \epa \delta_{k,pq} d\pp d\qq . 
\ea
The dummy variables $\pp$, $\qq$ and $s_p$, $s_q$, can be exchanged to symmetrise the equation; we find
\ba
\widehat{NL}(\kk) 
&=& \frac{1}{8} \sum_{s_p s_q} \int A_p^{s_p} A_q^{s_q} \frac{\epa \cdot (\ppnb \times \qqnb)}{\ppn \qpn}
(\ppn^2-\qpn^2)s_p s_q \epa \delta_{k,pq} d\pp d\qq \\
&&\mbox{} +\frac{i}{8} \sum_{s_p s_q} \int A_p^{s_p} A_q^{s_q} \frac{\epa \cdot (\ppnb \times \qqnb)}{\ppn \qpn}
\left( s_p \qpn - s_q \ppn \right)(\kknb \times \epa) \delta_{k,pq} d\pp d\qq . \nonumber
\ea
Coming back to the wave amplitude equation, we can write 
\ba
\left( \frac{\partial \hat \psi_k}{\partial t} - 2 i \ome \kpa \hat \phi_k \right) \kpn^2 \epa
+ \left( i \kpn^2 \frac{\partial \hat \phi_k}{\partial t} + 2 \ome \kpa \hat \psi_k \right) \kk \times \epa = \widehat{NL}(\kk) ,
\ea
therefore, after projection and use of the dispersion relation, we obtain
\begin{subequations}
\ba
\frac{\partial \hat \psi_k}{\partial t} - i \omega_k \kpn \hat \phi_k &=&
\sum_{s_p s_q} \int A_p^{s_p} A_q^{s_q} \frac{\epa \cdot (\ppnb \times \qqnb)}{8 \kpn^2\ppn \qpn} (\ppn^2-\qpn^2)s_p s_q \delta_{k,pq} d\pp d\qq , \quad \\
\frac{\partial \hat \phi_k}{\partial t} - i \omega_k \frac{\hat \psi_k}{\kpn} &=& 
\sum_{s_p s_q} \int A_p^{s_p} A_q^{s_q} \frac{\epa \cdot (\ppnb \times \qqnb)}{8 \kpn^2 \ppn \qpn} \left( s_p \qpn - s_q \ppn \right) \delta_{k,pq} d\pp d\qq . \quad
\ea
\end{subequations}
With the introduction of the canonical variables (\ref{vcanon}), the weighted addition of the previous expressions gives
\ba
\frac{\partial A^s_k}{\partial t} + i s \omega_k A^s_k &=& \sum_{s_p s_q} \int s s_p s_q  \frac{\epa \cdot (\ppnb \times \qqnb)}{8 \kpn \ppn \qpn} 
\left( \qpn^2-\ppn^2 + s \kpn \left( s_q \qpn - s_p \ppn \right) \right) \nonumber \\
&& \mbox{} \times A_p^{s_p} A_q^{s_q} \delta_{k,pq} d\pp d\qq  .
\ea
Remaking that
\be
\qpn^2-\ppn^2 = (s_q \qpn - s_p \ppn)(s_p\ppn + s_q \qpn) ,
\ee
we can rearrange the expression in the following manner
\ba
\frac{\partial A^s_k}{\partial t} + i s \omega_k A^s_k &=&
\sum_{s_p s_q} \int s s_p s_q  \frac{\epa \cdot (\ppnb \times \qqnb)}{8 \kpn \ppn \qpn} ( s_q \qpn - s_p \ppn) ( s \kpn + s_p \ppn + s_q \qpn) \nonumber \\
&& \mbox{} \times A_p^{s_p} A_q^{s_q} \delta_{k,pq} d\pp d\qq  .
\ea
We introduce the interaction representation for waves of weak amplitude ($0 < \epsilon \ll 1$)
\be
A_k^s = \epsilon a_k^s e^{-is\omega_k t} ,
\ee
and get eventually the wave amplitude equation after a few last manipulations
\be \label{WE}
\frac{\partial a^s_k}{\partial t} = \epsilon \sum_{s_p s_q} \int L_{kpq}^{ss_ps_q} a_p^{s_p} a_q^{s_q} e^{i \Omega_{k,pq} t} \delta_{k,pq} d\pp d\qq  ,
\ee
with $\Omega_{k,pq} \equiv s \omega_k - s_p \omega_p - s_q \omega_q$ and 
\be \label{coeff}
L_{kpq}^{ss_ps_q} \equiv  \omega_k \frac{\epa \cdot (\ppnb \times \qqnb)}{8 \kpn \ppn \qpn} s_p s_q \left( \frac{s_q \qpn - s_p \ppn}{s \omega_k}\right) 
( s \kpn + s_p \ppn + s_q \qpn) .
\ee
Expression (\ref{coeff}) satisfies the following properties (relation (\ref{reson}) is used)
\begin{subequations}
\ba
L_{0pq}^{ss_ps_q} &=& 0 , \\
L_{kqp}^{ss_qs_p} &=& L_{kpq}^{ss_ps_q} , \\
L_{pkq}^{s_pss_q} &=& - \frac{s_p \omega_p}{s \omega_k} L_{kpq}^{ss_ps_q} , \\
L_{kpq}^{-s-s_p-s_q} &=& - L_{kpq}^{ss_ps_q} , \\
L_{-kpq}^{ss_ps_q} &=& L_{kpq}^{ss_ps_q}  , \\
L_{-k-p-q}^{ss_ps_q} &=& L_{kpq}^{ss_ps_q} . 
\ea
\end{subequations}
The wave amplitude equation (\ref{WE}) governs the slow evolution of inertial waves of weak amplitude in the anisotropic limit. It is a quadratic non-linear equation which corresponds to the interactions between waves propagating along $\pp$ and $\qq$, in the positive ($s_{p},s_{q}>0$) or negative  ($s_{p},s_{q}<0$) direction. The multiple time scale method introduced in the next section is based on this expression. The symmetries listed above will also be used to simplify the derivation of the kinetic equation. Unlike the original derivation by \citet{Galtier2003}, expression (\ref{WE}) has been derived without going through a complex helicity basis. 
The wave amplitude equation tells us that the non-linear coupling between the states associated with the wavevectors $\ppnb$ and $\qqnb$ vanishes when these wavevectors are collinear. Moreover, we note that the non-linear coupling disappears when the wavenumbers $\ppn$ and $\qpn$ are equal if their associated directional polarities, $s_p$ and $s_q$, are also equal. 
These are general properties for helical waves \citep{Kraichnan1973,Waleffe1992,Turner2000,Galtier2003,Galtier2006,Galtier2014}.

\section{Asymptotic sequential closures}
\label{section3}

The method outlined here was first proposed by \citet{Benney1966} for three-wave interactions, but to our knowledge it has never been explicitly applied to a physical system. In fact, originally the main motivation for such a development was four-wave interactions that describe gravity waves \citep{Hasselmann1962,Newell1968}, and for which the main prediction of wave turbulence (Kolmogorov-Zakharov spectrum) is now well observed \citep{Lenain2017}. 

Unlike the classical perturbation analysis, the multiple time scale method is based on the existence of a sequence of time scales, $T_{0}$, $T_{1}$, $T_{2}$, ..., 
with by definition \citep{Nayfeh2004}
\be
T_{0} \equiv t, \quad T_{1}\equiv \epsilon t, \quad T_{2}\equiv \epsilon^{2} t, ... \, .
\ee
Because of the weak dependence in $t$ of $T_{1}$, $T_{2}$, ..., all these variables will be treated (it is an approximation) as independent. And the smaller $\epsilon$ is, the better the approximation.
The variation of the wave amplitude with $T_1$ and $T_2$ represents the slow variation that we wish to extract. (In practice, our analysis will be limited to $\mathcal{O} (\epsilon^2)$.) Using the chain rule, we obtain ($T_0$ being replaced by $t$)
\ba
\left({\partial \over \partial t} + \epsilon {\partial \over \partial T_{1}} + \epsilon^{2} {\partial \over \partial T_{2}} + ...\right) a^{s}_{k} 
=  \epsilon \sum_{s_{p} s_{q}} \int L^{s s_p s_q}_{kpq} a^{s_{p}}_{p} a^{s_{q}}_{q} e^{i \Omega_{k,pq}t} \delta_{k,pq} 
d\pp d\qq . && 
\label{eq6bis}
\ea
The variable $a^{s}_{k}$ must also be expanded to the power of $\epsilon$ and make the various scales appear in time
\be \label{adev}
a^{s}_{k} = \sum_{n=0}^{+\infty} \epsilon^n a^{s}_{k,n}(t,T_{1},T_{2},...) = a^{s}_{k,0} + \epsilon a^{s}_{k,1} + \epsilon^{2} a^{s}_{k,2} + ... \, .
\ee
Expression (\ref{adev}) is then introduced into the fundamental equation (\ref{eq6bis}); we obtain for the first three terms
\begin{subequations}
\begin{align}
{\partial a^{s}_{k,0} \over \partial t} &= 0 , \\
{\partial a^{s}_{k,1} \over \partial t} &= - {\partial a^{s}_{k,0} \over \partial T_{1}} + 
\sum_{s_{p} s_{q}} \int L^{s s_p s_q}_{kpq} a_{p,0}^{s_{p}} a_{q,0}^{s_{q}} e^{i \Omega_{k,pq}t} \delta_{k,pq} d\pp d\qq , \\
{\partial a^{s}_{k,2} \over \partial t} &= - {\partial a^{s}_{k,1} \over \partial T_{1}} - {\partial a^{s}_{k,0} \over \partial T_{2}} + 
\sum_{s_{p} s_{q}} \int L^{s s_p s_q}_{kpq} \left[ a_{p,1}^{s_{p}} a_{q,0}^{s_{q}} + a_{p,0}^{s_{p}} a_{q,1}^{s_{q}} \right] e^{i \Omega_{k,pq}t} \delta_{k,pq} d\pp d\qq . 
\end{align}
\end{subequations}
To lighten the writing, the time dependency of the variables has been omitted. After integration on $t$, one finds
\begin{subequations}
\begin{align}
a^{s}_{k,0} &= a^{s}_{k,0}(T_1,T_2,...) , \label{137} \\
a^{s}_{k,1} &= -t {\partial a^{s}_{k,0} \over \partial T_{1}} + b^{s}_{k,1}  , \label{138} \\
a^{s}_{k,2} &= \frac{t^2}{2} {\partial^2 a^{s}_{k,0} \over \partial T_{1}^2} -t {\partial a^{s}_{k,0} \over \partial T_{2}} 
- \int_0^t {\partial b^{s}_{k,1} \over \partial T_{1}} dt + \tilde b^{s}_{k,2}  , \label{139} 
\end{align}
\end{subequations}
with by definition
\be \label{bdelta}
b^{s}_{k,1} \equiv \sum_{s_{p} s_{q}} \int L^{s s_p s_q}_{kpq} a_{p,0}^{s_{p}} a_{q,0}^{s_{q}} \Delta(\Omega_{k,pq}) \delta_{k,pq} d\pp d\qq  , 
\ee
\ba
\Delta(X) \equiv \int_{0}^{t} e^{i Xt} dt  = \frac{e^{i Xt} -1}{i X} ,
\ea
and
\be \label{bdelta2a}
\tilde b^{s}_{k,2} \equiv \sum_{s_{p} s_{q}} \int L^{s s_p s_q}_{kpq}  \int_{0}^{t} 
\left[ a_{p,1}^{s_{p}} a_{q,0}^{s_{q}} + a_{p,0}^{s_{p}} a_{q,1}^{s_{q}}\right] e^{i \Omega_{k,pq}t} dt \delta_{k,pq} d\pp d\qq .
\ee
The previous equation is modified when expression (\ref{138}) is introduced; one finds
\ba
\tilde b^{s}_{k,2} &=& -\sum_{s_{p} s_{q}} \int L^{s s_p s_q}_{kpq}
\frac{\partial (a^{s_p}_{p,0} a_{q,0}^{s_{q}})}{\partial T_{1}} \left( \int_0^t t e^{i \Omega_{k,pq}t} dt \right) \delta_{k,pq} d\pp d\qq \\
&&\mbox{} + \sum_{s_{p} s_{q}} \int L^{s s_p s_q}_{kpq}
\left( \int_{0}^{t} (b^{s_p}_{p,1} a_{q,0}^{s_{q}} + b^{s_q}_{q,1} a_{p,0}^{s_{p}} ) e^{i \Omega_{k,pq}t} dt \right) 
\delta_{k,pq} d\pp d\qq . \nonumber
\ea
Expression (\ref{139}) becomes
\be \label{34e}
a^{s}_{k,2} = \frac{t^2}{2} {\partial^2 a^{s}_{k,0} \over \partial T_{1}^2} -t {\partial a^{s}_{k,0} \over \partial T_{2}} + b^{s}_{k,2} , 
\ee
with
\ba
b^{s}_{k,2} &=& -\sum_{s_{p} s_{q}} \int L^{s s_p s_q}_{kpq} \frac{\partial (a^{s_p}_{p,0} a_{q,0}^{s_{q}})}{\partial T_{1}} 
\left( \int_0^t \left[\Delta( \Omega_{k,pq}) + t e^{i \Omega_{k,pq}t} \right] dt \right)  \delta_{k,pq} d\pp d\qq \nonumber \\
&&\mbox{} + \sum_{s_{p} s_{q} s_{p'} s_{q'} } \int 2 L^{s s_p s_q}_{kpq} L^{s_p s_{p'} s_{q'}}_{pp'q'} a_{p',0}^{s_{p'}} a_{q',0}^{s_{q'}}  a_{q,0}^{s_{q}} 
\left( \int_{0}^{t} \Delta( \Omega_{p,p'q'}) e^{i \Omega_{k,pq}t} dt \right) \nonumber \\
&&\mbox{}  \times \delta_{k,pq} \delta_{p,p'q'} d\pp d\qq d\pp' d\qq' .
\ea
The time integrals give the relations
\ba
\int_0^t \left[\Delta( \Omega_{k,pq}) + t e^{i \Omega_{k,pq}t} \right] dt &=& t \Delta( \Omega_{k,pq}) , \\
\int_{0}^{t} \Delta( \Omega_{p,p'q'}) e^{i \Omega_{k,pq}t} dt &=& \frac{\Delta(\Omega_{k,p'q'q}) - \Delta(\Omega_{k,pq})}{i (\Omega_{k,p'q'q}-\Omega_{k,pq})} ,
\ea
that will be used below in the long time limit. 

\subsection{First asymptotic closure at time $T_1$}

With the previous definitions, the perturbative expansion of the second-order moment writes
\ba \label{37e}
\langle a^{s}_{k} a^{s'}_{k'} \rangle &=& 
\langle (a^{s}_{k,0} + \epsilon a^{s}_{k,1} + \epsilon^2 a^{s}_{k,2} + ...)(a^{s'}_{k',0} + \epsilon a^{s'}_{k',1} + \epsilon^2 a^{s'}_{k',2} + ...) \rangle \\
&=& \langle a^{s}_{k,0} a^{s'}_{k',0} \rangle + \epsilon \langle a^{s}_{k,0} a^{s'}_{k',1} + a^{s}_{k,1} a^{s'}_{k',0} \rangle 
+ \epsilon^2 \langle a^{s}_{k,0} a^{s'}_{k',2} + a^{s}_{k,1} a^{s'}_{k',1} + a^{s}_{k,2} a^{s'}_{k',0}  \rangle + ... , \nonumber 
\ea
where $\langle \rangle$ denotes the ensemble average.
We shall assume that this turbulence is statistically homogeneous. In this case, the second-order moment can be written in term of second-order cumulant, $q^{ss'}(\kk,\kk') \equiv q_{k}^{ss'}$, such that 
\be
\langle a^{s}_{k} a^{s'}_{k'} \rangle = q_{k}^{ss'} \delta (\kk+\kk') ,
\ee
where the presence of $\delta (\kk+\kk')$ is the consequence of the statistical homogeneity \citep{GaltierCUP2023}. 
We have to adapt the original formalism developed for isotropic problems \citep{Benney1966} to this anisotropic case where the dispersion relation depends not only on the wavenumber $k$ but also on the component $\kpa$. In this case, a contribution from $q^{ss'}$ is only relevant if $s=s'$, whereas it is for $s=-s'$ in the case of an isotropic problem. (With such conditions $q^{ss'}_{k}$ is real.)
We assume -- and this is the basic idea of the method -- that the second-order moment (in fact, the coefficient $q^{ss'}_{k}$ in front of the delta function) in the left hand side of expression (\ref{37e}) remains bounded at all time \citep{Benney1966}. 
As an example, we can think of the energy spectrum which, as we know, remains physically bounded. Therefore, the contributions on the right hand side must also be bounded at each order in $\epsilon$. We will see that secular terms can appear at different orders in $\epsilon$; this leads to certain conditions to cancel their contributions to keep the development uniform in time. As we shall see, at order $\mathcal{O} (\epsilon^2)$ this condition leads to the so-called kinetic equations. 
The main problem is therefore to account for all the secular contributions. 

At order $\mathcal{O} (\epsilon^0)$, we have the contribution of $\langle a^{s}_{k,0} a^{s'}_{k',0} \rangle$ which will therefore be assumed to be bounded at all times. 

At order $\mathcal{O} (\epsilon^1)$, we have the contribution
\ba \label{38e}
\langle a^{s}_{k,0} a^{s'}_{k',1} + a^{s}_{k,1} a^{s'}_{k',0} \rangle &=& 
\left\langle a^{s}_{k,0} \left(-t {\partial a^{s'}_{k',0} \over \partial T_{1}} + b^{s'}_{k',1}\right) + \left(-t {\partial a^{s}_{k,0} \over \partial T_{1}} + b^{s}_{k,1}\right) a^{s'}_{k',0} \right\rangle \nonumber \\
&=& -t \frac{\partial}{\partial T_{1}} \langle a^{s}_{k,0} a^{s'}_{k',0} \rangle + \langle a^{s}_{k,0} b^{s'}_{k',1} \rangle + \langle b^{s}_{k,1} a^{s'}_{k',0} \rangle .
\ea
The first term on the right hand side gives a secular contribution proportional to $t$. For the second term, we have
\be \label{39e}
\sum_{s_{p} s_{q}} \int L^{s s_p s_q}_{kpq}  \langle a^{s}_{k,0} a_{p,0}^{s_{p}} a_{q,0}^{s_{q}} \rangle \Delta(\Omega_{k,pq}) \delta_{k,pq} d\pp d\qq .
\ee
The long time limit ($t \gg 1/\omega$) of this oscillating integral will be given by the Riemann-Lebesgue lemma (the proof requires to use of generalised functions)
\ba
\Delta(X) &=& \frac{e^{i Xt} -1}{i X} \xrightarrow{\text{t $\to +\infty$}} \pi \delta(X) + i {\cal P} \left(\frac{1}{X}\right) , \label{RiemannL}
\ea
where ${\cal P}$ is the Cauchy principal value of the integral. Therefore, the long time limit of expression (\ref{39e}) gives no secular contribution. The same conclusion is obtained for the third term of expression (\ref{38e}). Also, the condition to cancel the unique secular term is
\be
\frac{\partial \langle a^{s}_{k,0} a^{s'}_{k',0} \rangle }{\partial T_{1}} = 0 ,
\ee
which means that the second-order moment does not evolve over a time scale $T_1$. As will be seen later, a turbulent cascade in inertial wave turbulence is only expected on a time scale $T_2$. (Here we have a point of disagreement with expression (2.43) in \citet{Benney1966}: it is not correct, but this has no impact on the rest of the paper.)

\subsection{Second asymptotic closure at time $T_2$}

The analysis continues at order $\mathcal{O} (\epsilon^2)$. With expression (\ref{34e}), the next contribution reads
\ba
&&\left\langle a^{s}_{k,1} a^{s'}_{k',1} + a^{s}_{k,0} a^{s'}_{k',2} + a^{s}_{k,2} a^{s'}_{k',0} \right\rangle = 
\left\langle \left(-t {\partial a^{s}_{k,0} \over \partial T_{1}} + b^{s}_{k,1}\right)\left(-t {\partial a^{s'}_{k',0} \over \partial T_{1}} + b^{s'}_{k',1}\right) \right\rangle \\
&& \mbox{} + \left\langle a^{s}_{k,0} \left(\frac{t^2}{2} {\partial^2 a^{s'}_{k',0} \over \partial T_{1}^2} -t {\partial a^{s'}_{k',0} \over \partial T_{2}} 
+ b^{s'}_{k',2} \right) + a^{s'}_{k',0} \left(\frac{t^2}{2} {\partial^2 a^{s}_{k,0} \over \partial T_{1}^2} -t {\partial a^{s}_{k,0} \over \partial T_{2}} + b^{s}_{k,2} \right) \right\rangle , \nonumber 
\ea
which gives after development and simplifications
\ba \label{eq44e}
\left\langle a^{s}_{k,1} a^{s'}_{k',1} + a^{s}_{k,0} a^{s'}_{k',2} + a^{s}_{k,2} a^{s'}_{k',0} \right\rangle &=& 
\frac{t^2}{2} \frac{\partial^2 \langle a^{s}_{k,0} a^{s'}_{k',0} \rangle }{\partial T_{1}^2} 
- t \frac{\partial \langle a^{s}_{k,0} a^{s'}_{k',0} \rangle }{\partial T_{2}} + \left\langle b^{s}_{k,1} b^{s'}_{k',1} \right\rangle \\
&-& \mbox{} t \left\langle {\partial a^{s}_{k,0} \over \partial T_{1}} b^{s'}_{k',1} + b^{s}_{k,1} {\partial a^{s'}_{k',0} \over \partial T_{1}} \right\rangle 
+ \left\langle a^{s}_{k,0} b^{s'}_{k',2} + a^{s'}_{k',0} b^{s}_{k,2} \right\rangle . \nonumber 
\ea
The first term on the right hand side cancels over the long time as required by the first asymptotic closure. The second term gives a secular contribution. The other three terms can potentially give a secular contribution: it is obvious for the fourth term and non-trivial for the third and fifth terms which require further development. 

The third term on the right hand side writes
\ba \label{46e}
\left\langle b^{s}_{k,1} b^{s'}_{k',1} \right\rangle &=& 
\sum_{s_{p} s_{q} s_{p'} s_{q'}} \int L^{s s_p s_q}_{kpq}  L^{s' s_{p'} s_{q'}}_{k'p'q'} 
\langle a_{p,0}^{s_{p}} a_{q,0}^{s_{q}} a_{p',0}^{s_{p'}} a_{q',0}^{s_{q'}} \rangle \Delta(\Omega_{k,pq}) \Delta(\Omega_{k',p'q'}) \nonumber \\
&&\mbox{} \times \delta_{k,pq} \delta_{k',p'q'} d\pp d\qq d\pp' d\qq' . 
\ea
Here again, the theory of generalised functions gives us the long time behaviour of this oscillating integral (with the Poincar\'e-Bertrand formula -- see e.g. \citet{Benney1969}) 
\ba
\Delta(X) \Delta(-X) \xrightarrow{\text{t $\to +\infty$}} 2 \pi t \delta(X) + 2 {\cal P} \left(\frac{1}{X}\right) \frac{\partial}{\partial X} .
\ea
Therefore, a secular contribution (proportional to $t$) involving a Dirac function is possible. The fourth-order moment in expression (\ref{46e}) can be decomposed into products of second-order cumulant plus a fourth-order cumulant such that (the statistical homogeneity is used as well as $\langle a_{k,0}^{s}\rangle = 0$)
\ba  \label{stat33}
&&\langle a_{p,0}^{s_{p}} a_{q,0}^{s_{q}} a_{p',0}^{s_{p'}} a_{q',0}^{s_{q'}} \rangle =  
q^{s_{p}s_{q}s_{p'}s_{q'}}_{pqp',0} \delta(\pp+\qq+\pp'+\qq') 
+ q^{s_{p} s_q}_{p,0} q^{s_{p'} s_{q'}}_{p',0} \delta(\pp+\qq) \delta(\pp'+\qq')  \nonumber \\
&&\mbox{} + q^{s_{p} s_{p'}}_{p,0} q^{s_{q} s_{q'}}_{q,0} \delta(\pp+\pp') \delta(\qq+\qq') 
+ q^{s_{p} s_{q'}}_{p,0} q^{s_{q} s_{p'}}_{q,0} \delta(\pp+\qq') \delta(\qq+\pp')  . 
\ea
Note that according to expression (\ref{37e}) by homogeneity we also have the relation $\kk=-\kk'$. We obtain 
\ba \label{48e}
&&\left\langle b^{s}_{k,1} b^{s_{k'}}_{k',1} \right\rangle = 
\sum_{s_{p} s_{q} s_{p'} s_{q'}} \int L^{s s_p s_q}_{kpq}  L^{s' s_{p'} s_{q'}}_{k'p'q'}
\left[ q^{s_{p}s_{q}s_{p'}s_{q'}}_{pqp',0} \delta(\pp+\qq+\pp'+\qq') \right.  \\
&&\mbox{}  + q^{s_{p} s_q}_{p,0} q^{s_{p'} s_{q'}}_{p',0} \delta(\pp+\qq) \delta(\pp'+\qq') 
+ q^{s_{p} s_{p'}}_{p,0} q^{s_{q} s_{q'}}_{q,0} \delta(\pp+\pp') \delta(\qq+\qq') \nonumber \\
&&\left. \mbox{} + q^{s_{p} s_{q'}}_{p,0} q^{s_{q} s_{p'}}_{q,0} \delta(\pp+\qq') \delta(\qq+\pp') \right] 
\Delta(\Omega_{k,pq}) \Delta(\Omega_{k',p'q'}) \delta_{k,pq} \delta_{k',p'q'} d\pp d\qq d\pp' d\qq' . \nonumber
\ea
We are looking for secular contributions. In the second line, the first term does not contribute since it imposes $\kk={\bf 0}$ which cancels $L^{s s_p s_q}_{kpq}$, but the second term can contribute (in this anisotropic problem) when the conditions $s_p=s_{p'}$ and $s_q=s_{q'}$ are satisfies. Likewise, in the third line a contribution is possible when $s_p=s_{q'}$ and $s_q=s_{p'}$. There is no contribution from the first line, which means that the situation is the same as if the distribution were Gaussian (however, we do not make this assumption). In summary, in the long time limit, the secular contribution, written ${\cal C}_t \left\langle b^{s}_{k,1} b^{s'}_{k',1} \right\rangle$, is 
\ba \label{49e}
{\cal C}_t  \left\langle b^{s}_{k,1} b^{s'}_{k',1} \right\rangle &=& 
4 \pi t \sum_{s_{p} s_{q}} \int L^{s s_p s_q}_{kpq} L^{s s_p s_q}_{k-p-q}
q^{s_{p} s_{p}}_{p,0} q^{s_{q} s_{q}}_{q,0} \delta (\Omega_{k,pq}) \delta_{k,pq} \delta_{kk'} d\pp d\qq \nonumber\\
&=& 4 \pi t \sum_{s_{p} s_{q}} \int \vert L^{s s_p s_q}_{kpq} \vert^2
q^{s_{p} s_{p}}_{p,0} q^{s_{q} s_{q}}_{q,0} \delta (\Omega_{k,pq}) \delta_{k,pq} \delta_{kk'} d\pp d\qq .
\ea

The fourth term on the right hand side of equation (\ref{eq44e}) does not contribute over the long time because it depends on the $T_1$ derivative. The proof is given by a new relation involving the n-order moments which can be written
\ba
\langle a_k^{s} a_{k'}^{s'} a_{k''}^{s''} ... \rangle &=& \langle a_{k,0}^{s} a_{k',0}^{s'} a_{k'',0}^{s''} ... \rangle \\
&&\mbox{} + \epsilon \langle a_{k,1}^{s} a_{k',0}^{s'} a_{k'',0}^{s''} ... + a_{k,0}^{s} a_{k',1}^{s'} a_{k'',0}^{s''} ... 
+ a_{k,0}^{s} a_{k',0}^{s'} a_{k'',1}^{s''} .... + ....\rangle \nonumber \\
&&\mbox{} + \epsilon^2 \langle .... \rangle + ... \nonumber
\ea
As for the second-order moment, we demand that the moments (ie. the coefficients in front of the delta functions) of order n are bounded. At order $\mathcal{O} (\epsilon)$, we obtain the relation
\ba
&&\langle a_{k,1}^{s} a_{k',0}^{s'} a_{k'',0}^{s''} ... + a_{k,0}^{s} a_{k',1}^{s'} a_{k'',0}^{s''} ... 
+ a_{k,0}^{s} a_{k',0}^{s'} a_{k'',1}^{s''} ... + ...\rangle \\
&& = \left\langle \left(-t {\partial a^{s}_{k,0} \over \partial T_{1}} + b^{s}_{k,1}\right) a_{k',0}^{s'} a_{k'',0}^{s''} ... 
+ a_{k,0}^{s} \left(-t {\partial a^{s'}_{k',0} \over \partial T_{1}} + b^{s'}_{k',1}\right) a_{k'',0}^{s''} ... + ... \right\rangle \nonumber \\
&& = -t {\partial \langle a^{s}_{k,0} a^{s'}_{k',0} a_{k'',0}^{s''} ... \rangle \over \partial T_{1}} 
+ \langle b^{s}_{k,1}a_{k',0}^{s'} a_{k'',0}^{s''} ... \rangle + \langle a^{s}_{k,0} b^{s'}_{k',1} a_{k'',0}^{s''} ... \rangle + ... \nonumber
\ea
Only the first term of the last line gives a secular contribution over the long time, which means that we have to impose the asymptotic condition
\be
{\partial \langle a^{s}_{k,0} a^{s'}_{k',0} a_{k'',0}^{s''} ... \rangle \over \partial T_{1}} = 0 ,
\ee
at any order n. Therefore, the probability density function does not depend on $T_1$ and we can assume that the variable itself does not depend on $T_1$.  
(It is a mild hypothesis because it is difficult to imagine a turbulent system where everything fluctuates, and in which it would be possible to have a $T_1$ dependence for the amplitude whereas the probability density function has no such dependence.)
This shows that the fourth term on the right hand side of equation (\ref{eq44e}) does not contribute in the long time. 

The last term of equation (\ref{eq44e}) writes 
\ba
&&\left\langle a^{s}_{k,0} b^{s'}_{k',2} + a^{s'}_{k',0} b^{s}_{k,2} \right\rangle = 
\sum_{s_{p} s_{q} s_{p'} s_{q'} } \int 2 L^{s' s_p s_q}_{k'pq} L^{s_p s_{p'} s_{q'}}_{pp'q'}
\left\langle a^{s}_{k,0} a_{p',0}^{s_{p'}} a_{q',0}^{s_{q'}}  a_{q,0}^{s_{q}} \right\rangle \nonumber \\
&&\mbox{} \times \left( \frac{\Delta(\Omega_{k',p'q'q}) - \Delta(\Omega_{k',pq})}{i (\Omega_{k',p'q'q}-\Omega_{k',pq})} \right) 
\delta_{k',pq} \delta_{p,p'q'} d\pp d\qq d\pp' d\qq' \nonumber \\
&&+\sum_{s_{p} s_{q} s_{p'} s_{q'} } \int 2 L^{s s_p s_q}_{kpq} L^{s_p s_{p'} s_{q'}}_{pp'q'}
\left\langle a^{s'}_{k',0} a_{p',0}^{s_{p'}} a_{q',0}^{s_{q'}}  a_{q,0}^{s_{q}} \right\rangle
\left( \frac{\Delta(\Omega_{k,p'q'q}) - \Delta(\Omega_{k,pq})}{i (\Omega_{k,p'q'q}-\Omega_{k,pq})} \right) \nonumber \\
&&\mbox{} \times \delta_{k,pq} \delta_{p,p'q'} d\pp d\qq d\pp' d\qq' .
\ea
The secular contributions will be given by the theory of generalised functions with the following relation
\be
\frac{\Delta(X) - \Delta(0)}{iX} \xrightarrow{\text{t $\to +\infty$}} \pi t \delta(X) + it {\cal P} \left(\frac{1}{X}\right)  .
\ee
We also need to use the following development (and its symmetric in $\kk'$)
\ba  \label{55e}
&&\langle a_{k,0}^{s} a_{p',0}^{s_{p'}} a_{q',0}^{s_{q'}} a_{q,0}^{s_{q}} \rangle =  
q^{ss_{p'}s_{q'}s_{q}}_{kp'q',0} \delta(\kk+\pp'+\qq'+\qq) 
+ q^{s s_{q}}_{k,0} q^{s_{p'} s_{q'}}_{p',0} \delta(\kk+\qq) \delta(\pp'+\qq') \nonumber \\
&&\mbox{} + q^{s s_{q'}}_{k,0} q^{s_{p'} s_{q}}_{p',0} \delta(\kk+\qq') \delta(\pp'+\qq) 
+ q^{s s_{p'}}_{k,0} q^{s_{q'} s_{q}}_{q',0} \delta(\kk+\pp') \delta(\qq'+\qq)  . 
\ea
In the right hand side of expression (\ref{55e}), the first two terms do not give a secular contribution, however, the last two terms give a contribution when the following conditions are satisfied, namely $s=s_{q'}$, $s_{p'}=s_q$ and $s=s_{p'}$, $s_{q'}=s_q$, respectively. After substitution and simplification, we obtain the secular contributions in the long time limit
\ba 
&&{\cal C}_t \left\langle a^{s}_{k,0} b^{s'}_{k',2} + a^{s'}_{k',0} b^{s}_{k,2} \right\rangle = \\
&&\mbox{}+4t \sum_{s_{p} s_{q}} \int L^{s' s_p s_q}_{k'pq} L^{s_p s_{q} s'}_{p-qk'} 
q^{s' s'}_{k',0} q^{s_{q} s_{q}}_{q,0}
\left( \pi \delta(\Omega_{k',pq}) + i {\cal P} \left(\frac{1}{\Omega_{k',pq}}\right) \right) 
\delta_{k',pq} \delta_{kk'} d\pp d\qq \nonumber \\
&& \mbox{}+4t \sum_{s_{p} s_{q}} \int L^{s s_p s_q}_{kpq} L^{s_p s_{q} s}_{p-qk} 
q^{s s}_{k,0} q^{s_{q} s_{q}}_{q,0}
\left( \pi \delta(\Omega_{k,pq}) + i {\cal P} \left(\frac{1}{\Omega_{k,pq}}\right) \right) 
\delta_{k,pq} \delta_{kk'} d\pp d\qq \nonumber \\
&&\mbox{}= 8 \pi t \sum_{s_{p} s_{q}} \int L^{s s_p s_q}_{kpq} L^{s_p s_{q} s}_{p-qk}
q^{s s}_{k,0} q^{s_{q} s_{q}}_{q,0}
\delta(\Omega_{k,pq}) \delta_{k,pq} \delta_{kk'} d\pp d\qq \nonumber .
\ea
The last writing is obtained by using the general property $L^{s s_p s_q}_{-k-p-q} = L^{s s_p s_q}_{kpq}$ and the symmetry in $\pp \to -\pp$ and $\qq \to -\qq$. 

If in expression (\ref{eq44e}) we impose the nullity of the sum of the different secular contributions, we find the asymptotic condition (after integration over $k'$)
\ba
\frac{\partial q^{ss'}_{k,0}}{\partial T_{2}} &=& 4 \pi \sum_{s_{p} s_{q}} \int \vert L^{s s_p s_q}_{kpq} \vert^2
q^{s_{p} s_{p}}_{p,0} q^{s_{q} s_{q}}_{q,0} \delta (\Omega_{k,pq}) \delta_{k,pq} d\pp d\qq \nonumber \\
&&\mbox{} + 8 \pi \sum_{s_{p} s_{q}} \int L^{s s_p s_q}_{kpq}  L^{s_p s_q s}_{p-qk} 
q^{s s}_{k,0} q^{s_{q} s_{q}}_{q,0} \delta(\Omega_{k,pq}) \delta_{k,pq} d\pp d\qq . \nonumber
\ea
Introducing $e^{s}_k \equiv q^{ss}_{k,0}$, we end up with (some simple manipulations are also used to symmetrise the equation)
\ba \label{KEIW}
\frac{\partial e_k^{s}}{\partial t} &=& \frac{\pi \epsilon^2}{16 s\omega_k} \sum_{s_{p} s_{q}} \int  
\left( \frac{\sin \theta_k}{\kpn} \right)^2 (s_p \ppn - s_q \qpn)^2 ( s \kpn + s_p \ppn + s_q \qpn)^2 \\
&&\mbox{} \times \left[ s\omega_k e_p^{s_p} e_q^{s_q} + s_p \omega_p e_k^{s} e_q^{s_q} + s_q \omega_q e_k^{s} e_p^{s_p} \right] 
\delta (\Omega_{kpq}) \delta_{kpq} d\pp d\qq ,  \nonumber
\ea
with $\theta_k$ the opposite angle to $\kknb$ in the triangle $\kknb=\ppnb+\qqnb$. Expression (\ref{KEIW}) is the kinetic equation for inertial wave turbulence in the anisotropic limit (see equation 7 in \citet{Galtier2003}; the small difference (sign and numerical factor) depends only on the normalisation of the canonical variables (\ref{vcanon})). This result proves that the anisotropic and asymptotic limits commute. Note that the kinetic equation for inertial wave turbulence does not describe the slow mode ($\kpa=0$) which involves strong turbulence.

\section{Discussion and conclusion}
\label{section4}

Our study completes the original derivation of \cite{Galtier2003} where the delicate issue of the asymptotic limit was mentioned but not explicitly used. We obtained the kinetic equation using a multiple time scale method that leads to a sequential asymptotic closures at times $T_1$ and $T_2$. This method consists of imposing the nullity of secular terms (proportional to $t$) which emerge over asymptotically long times in order to guarantee a bounded value for the associated moments and thus keep the development uniform in time. These secular terms do not involve fourth-order cumulants but only products of second-order cumulants. Consequently, the derivation of the kinetic equation (\ref{KEIW}) is performed in a systematic and consistent manner, and does not require any closure assumptions such as quasi-Gaussianity. In addition, the random phase approximation is natural as it arises dynamically from the separation of time scales. The latter property was first mentioned by \cite{Benney1966}, but has not always been recognized as an inherent property of wave turbulence \citep{ZLF92,Nazarenko11}. Generally, the random phase approximation is introduced (initially or subsequently) to fully justify the closure. This fundamental difference with the multiple time scale method suggests that the latter is the most natural method for deriving the kinetic equation of wave turbulence.
Note, however, that the derivation made, although systematic, says nothing about the remaining terms (not used to obtain the kinetic equation) in the small $\epsilon$ perturbation expansion, the implicit conjecture being that they are subdominant. The proof of this conjecture remains a mathematical challenge \citep{Deng2021}. 

The domain of validity of wave turbulence has already been discussed in \cite{Galtier2003}. The conclusion is that we can always find a finite domain where inertial wave turbulence exists. To show this, we introduce the linear time $\tau_W \sim 1/ \omega_k \sim k/(k_\parallel \Omega_0)$ and the non-linear time $\tau_{NL} \sim 1/(k u)$. In the anisotropic limit, we have $k \sim k_\perp$, and we obtain the time ratio
\begin{equation}
\chi = \frac{\tau_W}{\tau_{NL}} \sim \frac{k_\perp^2 u}{k_\parallel \Omega_0} ,
\end{equation} 
which depends on the scale. If the system under study is initially excited locally at large scale, isotropically, and with a Rossby number much smaller than $1$, then we can obtain $\chi \ll 1$ at large scale, which is the condition for having weak turbulence. As we have explained, such turbulence becomes anisotropic, with the energy cascading towards a region of Fourier space where $k_\perp \gg k_\parallel$. This leads to an increase of $\chi$ (the dependence of $u$ on wavenumbers does not alter this trend). Clearly, at sufficiently large values of $k_\perp$, we can find $\chi \sim 1$, which is synonymous with strong turbulence (and the critical balance regime). There is therefore a domain in Fourier space where the condition for the validity of wave turbulence can be satisfied even in the anisotropic limit.

In summary, it can be said that inertial wave turbulence for three-wave interactions is characterised by a dynamics on two time scales. On short time scales, of the order of the wave period, there is phase mixing which leads, due to the dispersive nature of the waves, to the decoupling of the correlations if there are initially present and to a statistics close to Gaussianity, as expected from the central limit theorem. This happens with a decay in $1/t$. 
On a longer time scale, the non-linear coupling -- weak at short times -- becomes non-negligible due to the resonance mechanism. This coupling leads to a regeneration of the cumulants via the product of lower order cumulants. It is these terms that are at the origin of the energy transfer mechanism. 

The second novelty of our study is the demonstration that anisotropy and asymptotic closure commute. (Note that this property was also found in Alfv\'en wave turbulence \citep{Galtier2002}.) Therefore, to obtain the main properties of inertial wave turbulence (exact power law solution, direction of the cascade, Kolmogorov constant, existence of an inertial range), the limit $\kpn \gg \kpa$ can be taken before any statistical development. The study also reveals that it is not necessary to use an helicity basis,  which simplifies the treatment. The third novelty is that the multiple time scale method has been generalized to an anisotropic problem involving different types of correlation in terms of directional polarity. 

Note that in the present derivation, the system under study is assumed to be of infinite size and can therefore be treated as continuous. The numerical simulation with its grid of points escapes this description. Effects (freezing of the cascade) linked to the discretisation of the Fourier space can appear because the resonance conditions are a priori more difficult to satisfy (see for example \citet{Connaughton2001} for capillary waves). In theory, the weaker the non-linearities, the more important these effects are. In inertial wave turbulence, \citet{Bouroudia2008} has shown that discretisation effects become non-negligible when the Rossby number, $R_{o}$, is smaller than $10^{-3}$. Beyond this value, but still for a small $R_{o} \ll 1$, these effects are negligible because of the quasi-resonances which, with the resonances, contribute to the energy transfer.

\acknowledgements
I would like to thank and pay tribute to Professor Vladimir Zakharov, whose contribution to wave turbulence has been profound and inspiring for me.

\bibliography{WT-Biblio}

\end{document}